\documentclass[josab,osajnl,twocolumn,showpacs,superscriptaddress,10pt]{revtex4-1} 
\usepackage{amsmath,amssymb,graphicx}
\usepackage{subfigure}
\usepackage{palatino}
\usepackage{hyperref}

\begin{document}

\title{\rm  \bfseries Quantifying the source of enhancement in experimental continuous variable quantum illumination}

\author{\rm Sammy Ragy}
\email{pmxsr3@nottingham.ac.uk}
\affiliation{School of Mathematical Sciences, The University of Nottingham, University Park, Nottingham NG7 2RD, United Kingdom}

\author{\rm Ivano Ruo Berchera}
\affiliation{INRIM, Strada delle Cacce 91, I-10135 Torino, Italy}

\author{\rm Ivo P.~Degiovanni}
\affiliation{INRIM, Strada delle Cacce 91, I-10135 Torino, Italy}

\author{\rm Stefano Olivares}
\affiliation{Dipartimento di Fisica, Universit\`a degli Studi di Milano, and CNISM UdR Milano Statale, I-20133 Milano, Italy}

\author{\rm Matteo G.~A.~Paris}
\affiliation{Dipartimento di Fisica, Universit\`a degli Studi di Milano, and CNISM UdR Milano Statale, I-20133 Milano, Italy}

\author{\rm Gerardo Adesso}
\affiliation{School of Mathematical Sciences, The University of Nottingham, University Park, Nottingham NG7 2RD, United Kingdom}

\author{\rm  Marco Genovese}
\affiliation{INRIM, Strada delle Cacce 91, I-10135 Torino, Italy}

\begin{abstract}
A quantum illumination protocol exploits correlated light beams to enhance the probability of detection of a partially reflecting object lying in a very noisy background. Recently a simple photon-number-detection based implementation of a quantum illumination-like scheme has been provided in [Lopaeva {\it et al,}, Phys. Rev. Lett. {\bf 101}, 153603 (2013)] where the enhancement is preserved despite the loss of non-classicality.  In the present  paper we investigate the source for quantum advantage in that realization.  We introduce an effective two-mode description of the light sources and analyze the mutual information as quantifier of total correlations in the effective two-mode picture. In the relevant regime of a highly thermalized background, we find that the improvement in the signal-to-noise ratio achieved by the entangled sources over the unentangled thermal ones amounts exactly to the ratio of the effective mutual informations of the corresponding sources. More precisely, both quantities tend to a common limit specified by the squared ratio of the respective  cross-correlations. A thorough analysis of the experimental data confirms this theoretical result.
\end{abstract}

\maketitle

\section{Introduction}

Quantum illumination is a scheme for target detection embedded in very noisy environments which provides improvement when using entangled input probes over any possible
separable state \cite{Lloyd,Tan}. This is notable since the scheme revolves around the presence of realistic noise and loss, which are typically the enemies of quantum enhanced
schemes \cite{q:based:a,q:based:b}. The continuous variable version of quantum illumination \cite{Tan}
is based on states with Gaussian Wigner function \cite{ReviewW,ReviewO,ReviewG} such as the so-called twin beams, which are routinely produced by parametric processes or by the interference of squeezed states \cite{ou:92,laurat:05,porzio:09,schna:13,oli:fid}. The noisy environment is mimicked by an incoherent, thermal background \cite{Tan}. Surprisingly, the advantage appears even though the entanglement
is completely compromised by the noise, posing the question of where the quantum enhancement comes from.
\par
In its most general formulation, quantum illumination considers rigorous discrimination theory based on the Chernoff bound \cite{Lloyd,Tan}. This allows the distinction between two hypotheses (target present or target absent), represented by different probability distributions (or density matrices in quantum theory) \cite{Chernoff,ChernoffG}. However, no detector has yet been conceived of saturating the bound and even suboptimal detectors, though theoretically proposed, have not been experimentally realized \cite{Guha}.
\par
Recently, a pragmatic implementation of a continuous variable quantum illumination-like scheme was experimentally achieved by Lopaeva {\it et al.} \cite{Lopaeva}.
This particular scheme is realized as an `intensity-interferometry' type experiment, wherein the output intensities of two detectors are correlated,
yielding a fourth-order correlation in the field operators.
The noise is introduced by a pseudo-thermal field directly addressed to the detector in the object arm, whose photons are regarded as indistinguishable from the ones
of the twin beam if only one detector output is considered; that is, they only differ in the nature of their correlations. This experiment, aimed to demonstrate quantum enhancement in target detection in a realistic situation,
presents some differences with the previous theoretical schemes.
The noise modes additively increase the number of detected photons, without interfering with any light reflected by the target.
Thus, the detected field has a multimode structure and cannot be represented by a single Gaussian mode in the usual form.
This prevents us from gaining much insight from previous theoretical analysis and renders entanglement at the detector difficult to evaluate. On the other hand, by using quantumness criteria related to the Glauber Sudarshan $P$-representation, it is possible to show that the quantum scheme is largely more powerful than any classical scheme with the same local statistical properties, in terms of photon number detection and correlation measurements. As in the case of \cite{Lloyd,Tan}, in the the scheme of \cite{Lopaeva}, entanglement is completely destroyed before the detection stage.
\par
The interesting question that naturally arises is about the actual source of the quantum enhancement. This was discussed in the original exposition of Gaussian quantum illumination \cite{Tan} (see also the very recent further experiment in \cite{Zhang}): if we take identical single-mode statistics, then the only difference between using entangled light and unentangled light is the maximum allowable magnitude of the cross-correlations. Therefore, these are expected to yield the enhancement. This question has also recently been addressed  for a discrete variable setting, in which case it was suggested that quantum discord can account for the resilience of quantum illumination \cite{vedral:13}. Quantum correlations of the discord type have also been investigated as resources for parameter estimation \cite{blind,GIP} and channel discrimination \cite{Giovannetti}, including quantum illumination settings, in worst case scenarios.

\par
In this paper, we focus on the continuous variable system considered in \cite{Lopaeva}  and we approach the problem from the view-point of information analysis.
As mentioned above, the main obstacle to pursuing this goal is the multimode nature of the involved states, which prevents direct use of the most basic tools of Gaussian quantum information \cite{ReviewW,ReviewO,ReviewG}. Nevertheless, it has been observed in \cite{Ghost}
that for interferometric
setups based on intensity measurements on bipartite multimode Gaussian
states, e.g.~ghost imaging, it is possible under particular assumptions to
introduce an effective two-mode description, which is useful to obtain
theoretical predictions for some particular quantities of interest. In particular, it was pointed out that for separable light modes
the signal-to-noise ratio (SNR) is is closely related to mutual information (MI), which represents total correlations in the effective two-mode picture.

Here, we adopt methods similar to those of \cite{Ghost} in order to achieve a novel and simple quantitative investigation of the quantum enhancement obtained in the experimental demonstration of Ref~\cite{Lopaeva}. Ultimately, we show that the correspondence between SNR and MI holds also for this scheme. More precisely, we demonstrate that the ratio between the SNRs obtained with entangled and unentangled light, respectively, which was measured in the experiment \cite{Lopaeva} and can be regarded as an indicator of the quantum enhancement, converges to the ratio between the respective MIs. While specialized to a particular implementation, this result facilitates interpretation and understanding of the phenomenon of quantum illumination.

The paper is structured as follows. In Section~\ref{s:setup} we illustrate the setup of the considered quantum illumination protocol and introduce the effective description to deal with the presence of a very large number of modes and a high level of thermal noise. Section~\ref{s:theo} reports on the theoretical results, which are then compared with the experimental ones in Section~\ref{s:exp}. We conclude the paper with a discussion in Section~\ref{s:concl}.

\section{Setup and analysis}\label{s:setup}
\begin{figure}[t]
\includegraphics[width=8.5cm]{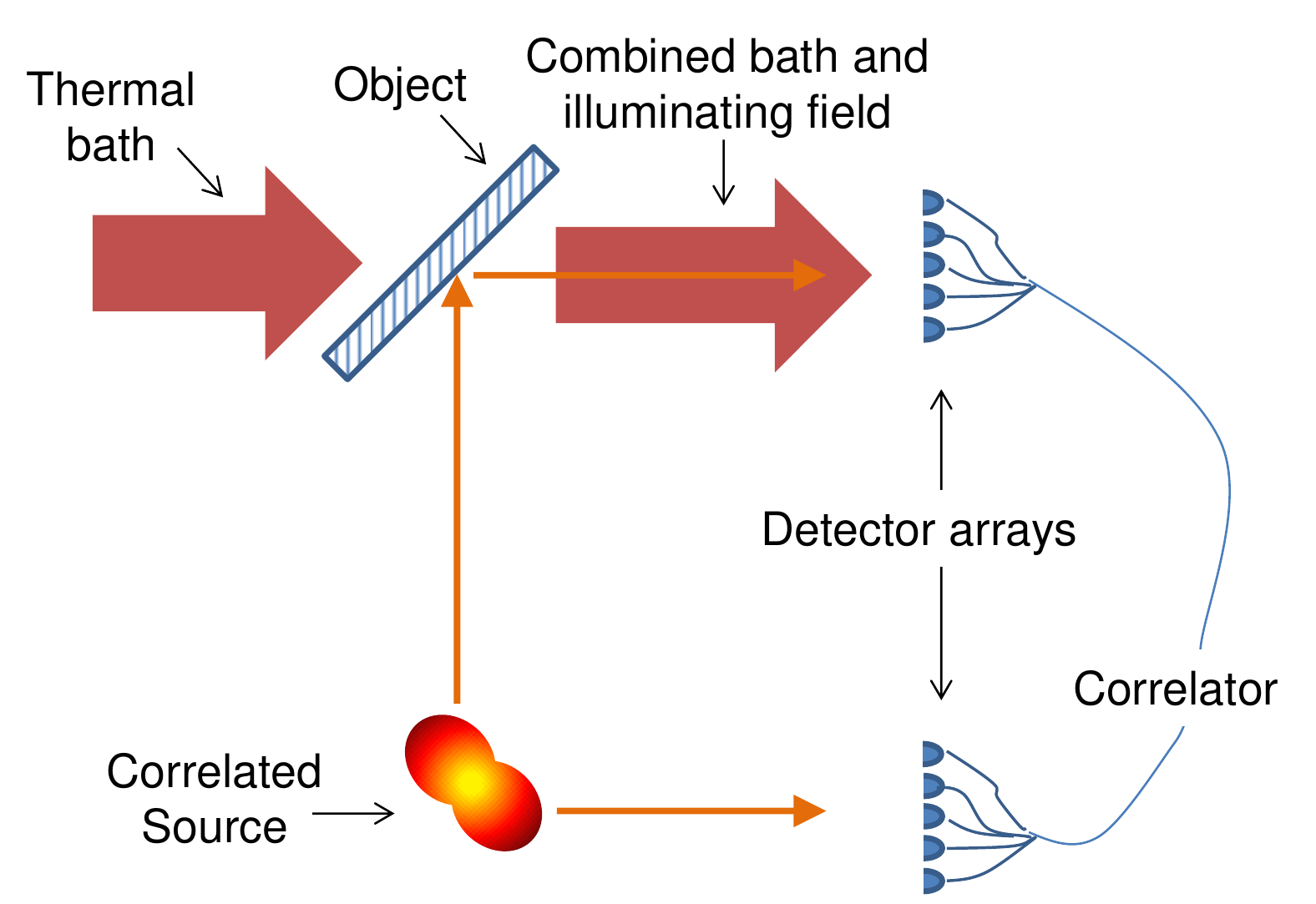}
\caption{A schematic diagram of a quantum illumination scheme. Here the object under consideration is a beam splitter. A portion of the light produced by the correlated source will be reflected by the beam splitter and the challenge is to detect this beneath the dominant thermal noise, thereby discriminating the presence of the object.}
\label{fig:Quillu} \end{figure}

In the experimental setup implemented in Ref.~\cite{Lopaeva}  the object to detect is a 50:50 beam splitter (BS) embedded in a ``bath''  of thermal modes. The light source used to probe the presence of the object consists of multiple identical and independent pairs of either twin beams (TWB) or classical-correlated thermal beams (THB). Charge-coupled device (CCD) arrays are placed in each of the signal and reference planes, as outlined in the scheme of Fig.~\ref{fig:Quillu}. Each pixel in the signal plane collects $M$ TWB (or THB) modes resulting in a net photon count per pixel $N_{I}$ (excluding the bath), which are correlated with $M$ corresponding modes intercepting another pixel in the reference plane counting $N_{r}$ photons (see also \cite{SSNQI,BridaPRL2009}). In our analysis, we dub the mean value of the count of the illuminating field as $N=\langle N_{r}\rangle$, while $N_\beta$ and  $M_\beta$ are the  photon count and the number of modes per pixel of the bath. The losses are taken into account by the detection efficiencies $\eta$ for the illuminating light (where this quantity does not include the non-unit reflectivity of the object in the reference plane) and $\eta_\beta$ for the bath.
\par
The covariance of photon counts per pixel on the signal plane $N_{s}=N_I+N_\beta$ and on the reference plane $N_{r}$ is evaluated averaging over the set of pixel pairs $N_{pix}$ in one shot of the CCD ($N_{pix}=80$ in the experiment). The covariance is expected to vanish when the object is absent, so the detection of the target is declared when the covariance is larger than a certain threshold value. The SNR (normalized by $N_{pix}^{1/2}$) for the measurement, given by the difference of covariances in the cases when the object is present (in) or absent (out), is~\cite{Lopaeva}
\begin{equation}\label{sonar}
\text{SNR}=\frac{|\langle\Delta^{\text{in}}-\Delta^{\text{out}}\rangle|}{\sqrt{\delta^2\Delta^\text{in}+\delta^2\Delta^\text{out}}}
\end{equation}
for $\Delta= N_ s N_r-\langle N_s \rangle \langle N_r\rangle$, where $\delta^2 X= \langle X^{2}\rangle- \langle X\rangle^{2}$. Remarkably, the TWB entangled input, representing the maximal allowable cross-correlation between two modes, harbours a hefty improvement over the case of maximally correlated modes within the bounds of separability i.e. with a proper $P$-representation, as experimentally demonstrated in  \cite{Tan,Lopaeva}.
\par
The complete experimental setup we are investigating is extremely complicated due to the presence of many modes and high levels of noise. This makes it unfeasible to pursue a direct analytical approach to the problem. We can overcome this difficulty by invoking a practical, effective description, which takes into account both the multimodal nature of the involved fields and the `coarse-graining' occurring at the detection stage, in which each CCD pixel collect a quite high number of modes. Fortunately, since the system involves Gaussian states, we can consider the approach introduced in \cite{Ghost}, which allows us to reduce the description to a simple two-mode effective system. Considering the nature of the involved states, for each pixel we can introduce the following effective operators
\begin{equation}\label{effing}
\hat{a}_\text{eff} = \frac{1}{\sqrt{M+M_\beta}}\sum_{k=1}^{M+M_\beta} \hat{a}_k,
\end{equation}
where $\hat{a}_k$ refer to the individual mode operators in our system. The effective modes are therefore a linear combination of the original ones, respecting the canonical commutation relations $[\hat{a}_\text{eff} ,\hat{a}_\text{eff} ^{\dag}]=1$.
An intuitive, operational picture of this averaging can be obtained by considering the combination of all $M+M_\beta$ modes equally by using a series of $M+M_\beta$ beam splitters with appropriate transmissivities. Whilst it is entirely impractical to actually perform this step in an experimental setting, we can still deduce an equivalent quantity with the given experimental set-up. Moreover, since the detectors in the actual experiment do not resolve individual modes, this coarse-graining provides a feasible average of what the detector `sees': for example, if all the modes are identical and pairwise correlated with a mode in the opposite plane, then the effective operators for each detector give identical correlations to those of the individual mode pairs impinging upon the detectors \cite{Ghost}.

We take the propagation of light from the source to the far-field enacted by a lens, such that we achieve a one-to-one correspondence between the transverse momentum-parametrized modes in the source plane and the transverse position-parametrized modes in the detection planes. Nevertheless, by analogy with previous works on the topic, the results of our analysis are reproducible for near-field propagation as well \cite{Ghost, Gatti}.
By merit of the far-field propagation we have $\langle \hat{a}^\dagger_k \hat{a}_{k'} \rangle=\delta_{k,k'} \langle \hat{a}^\dagger_k \hat{a}_{k'} \rangle$ and the coarse-graining in Eq.~(\ref{effing}) establishes the arithmetic mean of the second order auto- and cross-correlations \cite{notesammy}.
\par
As noted, the effective modes are a linear combination of the original ones, hence informational quantities which depend on the first and second moments of the mode operators will be employed. In this case, we can further simplify our description by resorting to the established tools of bosonic Gaussian quantum information \cite{Review,ReviewW,ReviewO,ReviewG}. In fact, we can construct  effective two-mode covariance matrices of the quadrature operators, with elements $$\sigma_{kj}=\text{Tr}[\hat{\rho} (\hat{R}_k\hat{R}_j+\hat{R}_j\hat{R}_k)].$$ Here $\hat{R}_{k}$, $k=1,2,3,4$, are the elements of the quadrature vector given by $\hat{R}=(\hat{q}_{\text{eff},s},\hat{p}_{\text{eff},s},\hat{q}_{\text{eff},r},\hat{p}_{\text{eff},r})^T$, where $\hat{a}_\text{eff} = (\hat{q}_\text{eff} + i \hat{p}_\text{eff})/\sqrt{2}$. From these covariance matrices we can fully extract any relevant informational measure based on the first and second moments.

 In particular, here we consider the MI calculated via the R\'enyi entropy  $S_2$ \cite{arenyi} as an informational quantity to compare with the SNR.  R\'enyi MI is a measure of total correlations which, for Gaussian states, is equally valid as the conventional one defined in terms of  von Neumann entropy, and which enjoys a clear operational interpretation in terms of phase-space sampling of the two-mode Wigner function by homodyne detections \cite{Renyi}. Precisely, the R\'enyi MI measures the extra information (in natural bits) that needs to be transmitted over a continuous variable channel to reconstruct the complete joint Wigner function of a two-mode state, rather than the sole marginal Wigner functions of each of the two subsystems separately. In this respect, it is an intuitive measure of total quadrature correlations between the two modes. We further observe that the R\'enyi MI is simpler to compute than the von Neumann MI, which makes it a convenient choice for our continuous variable setup, at variance with qubit scenarios where von Neumann entropic measures of correlations can be efficiently evaluated \cite{vedral:13}.

\section{Theoretical results}\label{s:theo}

For a Gaussian state $\rho$ with covariance matrix $\boldsymbol{\sigma}$, the R\'enyi entropy of order $2$ is simply given by $S_2(\rho) = \frac12 \ln (\det \boldsymbol{\sigma})$. Our quantity of interest is then given by the total correlations between the two effective modes, quantified by the (R\'enyi) MI, defined as \cite{Renyi}
\begin{equation}
\text{MI} (r:s)=S_2(\rho_r)+S_2(\rho_s)-S_2(\rho_{rs})\,.\end{equation}
The effective two-mode covariance matrices $\boldsymbol{\sigma}_{\text{THB}, \text{TWB}}$ when the illuminated object is a balanced beam splitter and the light source S consists of THB or TWB, respectively, take the standard form
\begin{equation*}
\boldsymbol{\sigma}_{\text{S}}=\left(
\begin{array}{cccc}
 a_{\text{S}} & 0 & c_{\text{S}} & 0 \\
 0 & a_{\text{S}} & 0 & d_{\text{S}} \\
 c_{\text{S}} & 0 & b_{\text{S}} & 0 \\
 0 & d_{\text{S}} & 0 & b_{\text{S}} \\
\end{array}
\right),
\end{equation*}
with ${\text{S}}= {\text{THB}},{\text{TWB}}$, and:
\begin{subequations}
\label{CMs}
\begin{align}
a_{\text{THB}}&=a_{\text{TWB}}=1+2 \eta \mu_1\,,\\[1ex]
b_{\text{THB}}&=b_{\text{TWB}}=1+\frac{\eta \mu_1M  +2 \eta_\beta \mu_\beta M_\beta }{M+M_\beta}\,,\\[1ex]
c_{\text{THB}}&=d_{\text{THB}}=\eta \mu_1 \sqrt{\frac{2 M}{M+M_\beta}}\,,\\[1ex]
c_{\text{TWB}}&=-d_{\text{TWB}}=\eta \sqrt{\mu_1^2 +\mu_1}
\sqrt{\frac{ 2M}{M+M_\beta}}\,,
\end{align}
\end{subequations}
where $\mu_1=N/(\eta M)$ is the mean photon count per mode at the source and $\mu_\beta=N_\beta /(\eta_\beta M_\beta)$ is the analogous quantity with respect to values for the bath.
\par
\begin{figure}[t]
\centering
\includegraphics[width=8.5cm]{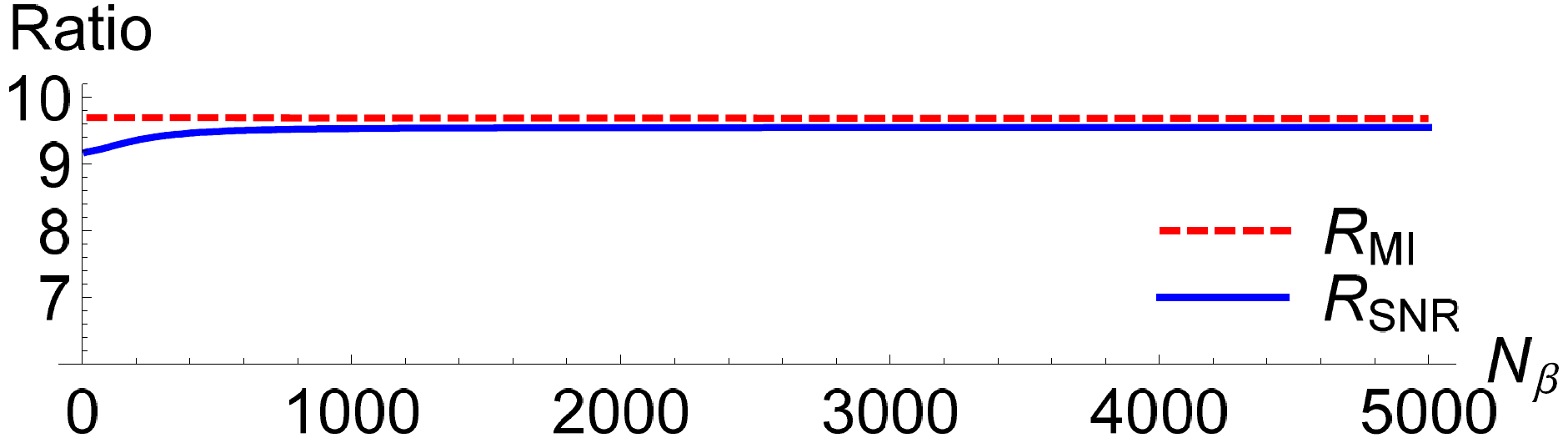}
\caption{(Color online) A theoretical plot of the ratios $R_{\text{SNR}}$ (blue solid) and $R_{\text{MI}}$ (red dashed) for quantum illumination with parameters set at realistic experimental values of $N=4000$, $M=90000$, $M_\beta=50$, $\eta=0.38$ and $\eta_\beta=0.5$. The asymptotic limit for $N_\beta \to \infty$ is given by Eq.~(\ref{cross}).
}
\label{PlotBath}
\end{figure}
Adopting the SNR (\ref{sonar}) as figure of merit,  we can quantify the enhancement achieved by the TWB over the THB  by considering the ratio of the respective SNRs \cite{Lopaeva},
\begin{equation}R_{\text{SNR}}=\frac{\text{SNR}_{\text{TWB}}}{\text{SNR}_{\text{THB}}}\,,\label{R_SNR}\end{equation}
for identical single-mode statistics. Similarly, we can analyse the `enhancement' in total effective correlations by defining the corresponding ratio of the MIs,
\begin{equation}
R_{\text{MI}}=\frac{\text{MI}_{\text{TWB}}}{\text{MI}_{\text{THB}}}\,.
\end{equation}

It is instructive to illustrate our findings by first plotting theoretical expectations for the comparisons of the respective ratios $R_{\text{SNR}}$ and $R_{\text{MI}}$. In Fig.~\ref{PlotBath}, we keep all parameters constant apart from the bath photon-count $N_\beta$. We notice that in the regime of highly thermal bath, which is the relevant regime for which the phenomenon of quantum illumination was defined \cite{Lloyd,Tan},  the ratios of SNR and MI converge to each other and become asymptotically identical. This is observed in all useful parameter regimes.
If one fixes indeed $N_\beta$ to a sufficiently large number which ensures a dominant bath (e.g.~$N_\beta=5000$), and lets the photon number $N$ of the illuminating field vary in a broad yet realistic regime (say from $10^2$ to $10^4$) an essentially perfect identity between  $R_{\text{MI}}$ and $R_{\text{SNR}}$ is retrieved.
\par
As one can intuitively expect, the common value attained by both the SNR and MI ratios in practical parameter regimes ($N_\beta \gg 1$) for quantum illumination is determined exactly by the cross-correlations squared, namely:
\begin{equation}\label{cross}
\lim_{N_\beta \rightarrow \infty} R_{\text{SNR}} = \lim_{N_\beta \rightarrow \infty} R_{\text{MI}} = \left\vert \frac{c_{\text{TWB}}}{c_{\text{THB}}} \right\vert^2\,,
\end{equation}
where the effective correlation elements $c_{\text{TWB},\text{THB}}$ are defined by Eq.~(\ref{CMs}). The proof of this for the case of the MI ratio follows by a generalization of the appendix in \cite{HBT}. For what concerns the SNR ratio, when the bath is dominant, the noise terms in the denominator of the SNRs are effectively independent of the source, TWB or TH, actually considered in the protocol and cancel each other, provided they have the same single beam fluctuation (see \cite{Lopaeva} for details). Thus:
 \begin{equation}
 R_\text{SNR}\approx \frac{|\langle\Delta^{\text{in}}-\Delta^{\text{out}}\rangle|_\text{TWB}}{|\langle\Delta^{\text{in}}-\Delta^{\text{out}}\rangle|_\text{THB}}\,.
 \end{equation}
It can be shown through the Gaussian moment factoring that this equates the ratio of cross-correlations squared, Eq.~(\ref{cross}).
\par
Interestingly, this finding provides a link between recent pieces of work which examined the role of various kinds of correlations in intensity interferometry schemes: in particular, previous studies compared either the SNR to the MI of effective operators \cite{Ghost} or the MI to the ratio of intensity covariances (though in quite a different context) \cite{HBT}. Within the quantum illumination setting considered here, in which all of these quantifiers can be defined and jointly analyzed, we find that they are all quantitatively connected.

In \cite{Lopaeva}, the SNR enhancement was also linked to the ratio of the generalized Cauchy-Schwartz parameter $\varepsilon\equiv\langle: \delta N_{s}\delta N_{r}:\rangle/(\langle :\delta^{2} N_{s}:\rangle \langle: \delta^{2} N_{r}:\rangle)^{1/2}$, where $\langle : \: :\rangle$ is the normally ordered quantum expectation value and $\varepsilon\leq1$ indicates a classical regime, i.e.~corresponding to a state having positive well-defined $P$-representation. In particular it has been shown that the quantum enhancement with respect to the optimal classical strategy, in the limit of dominant bath, is $R_{\text{SNR}}=\frac{\varepsilon_{\text{TWB}}}{\varepsilon_{\text{THB}}}>1$. Remarkably, in the presence of the bath one finds $\varepsilon_{\text{TWB}} \leq 1$ indicating classicality although the enhancement survives. Lastly, we mention that the same transition to the classical regime without affecting the performance improvement has been observed in \cite{P.Sc.Lopaeva} for yet another common parameter of non-classicality, the noise reduction factor $\langle \delta^{2}( N_{s}-N_{r})\rangle/(\langle N_{s}+ N_{r}\rangle)$ \cite{Bondani,Ishkakov,SSNQI, BridaPRL2009}.

This series of observations shows how different parameters, originally introduced to assess experimental quality, are in fact all capturing the same physics in the case of Gaussian light sources for quantum illumination, and are thus all able to reveal the quantum advantage of the scheme, even in a regime in which quantumness in the form of entanglement appears not to manifestly survive.

\begin{figure}[b!]
\subfigure{\includegraphics[width=8.5cm]{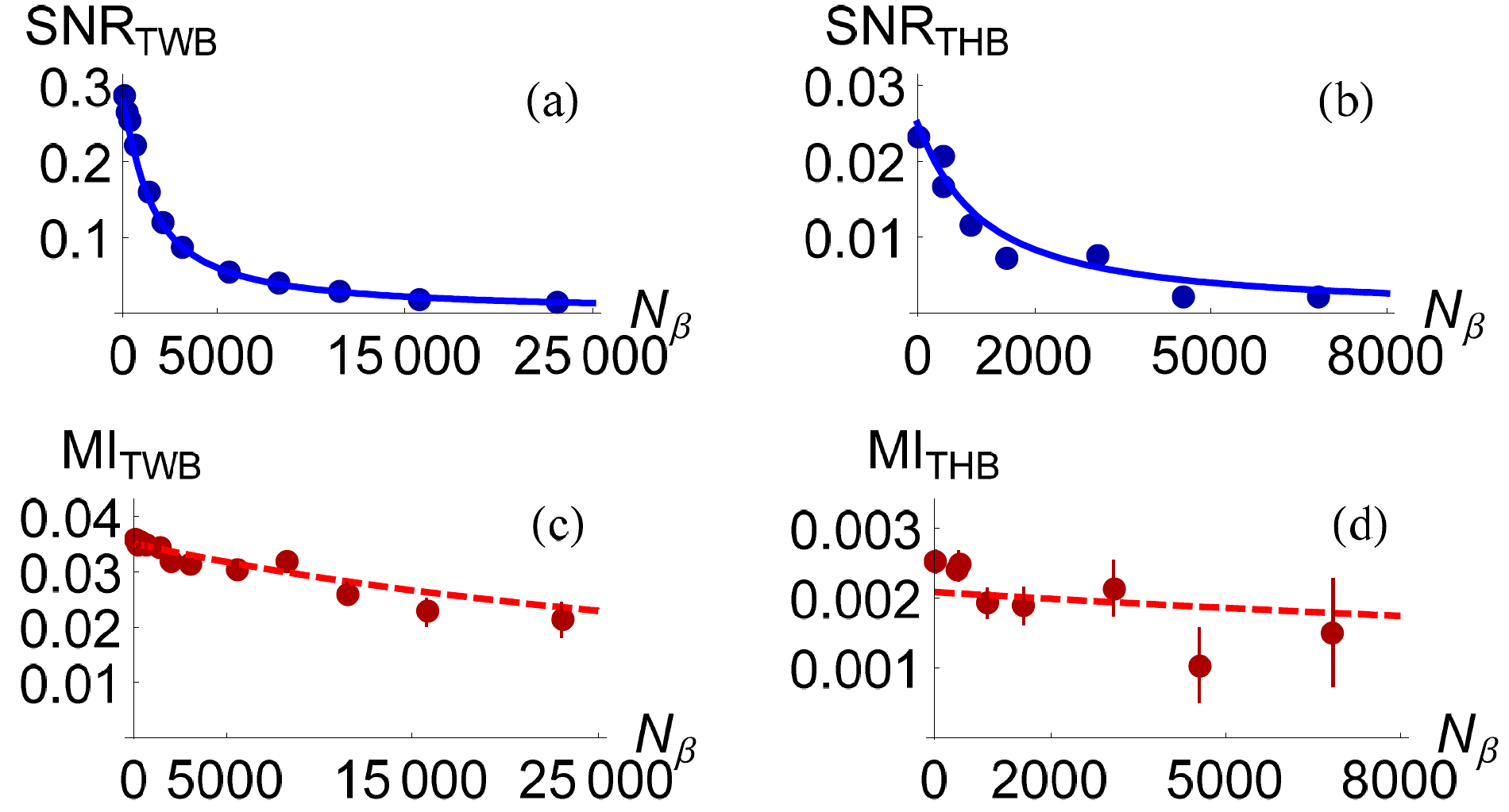}}\\
\setcounter{subfigure}{4}
\subfigure[ ]{\includegraphics[width=8.8cm]{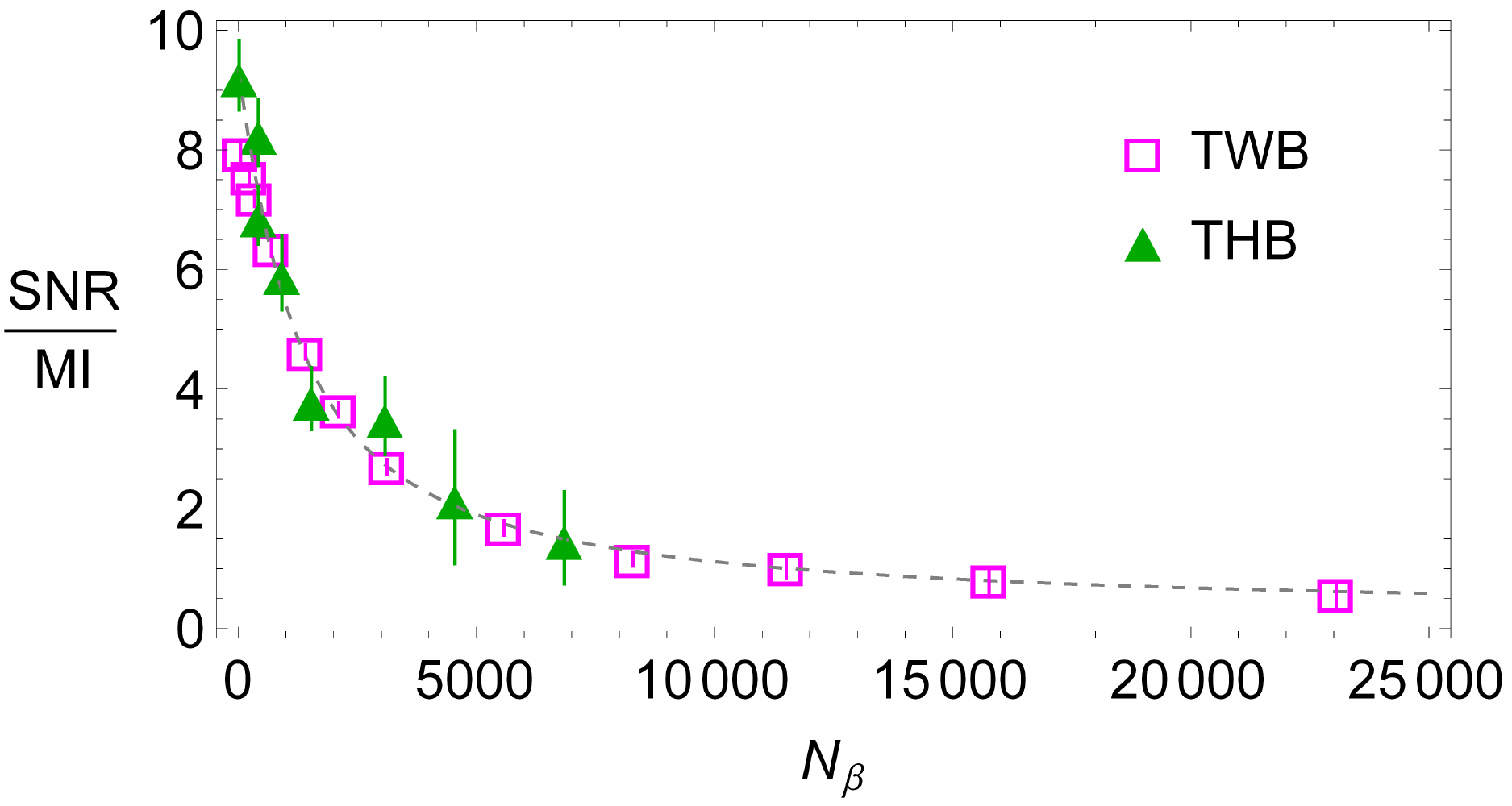}}\\
\subfigure[ ]{\includegraphics[width=8.8cm]{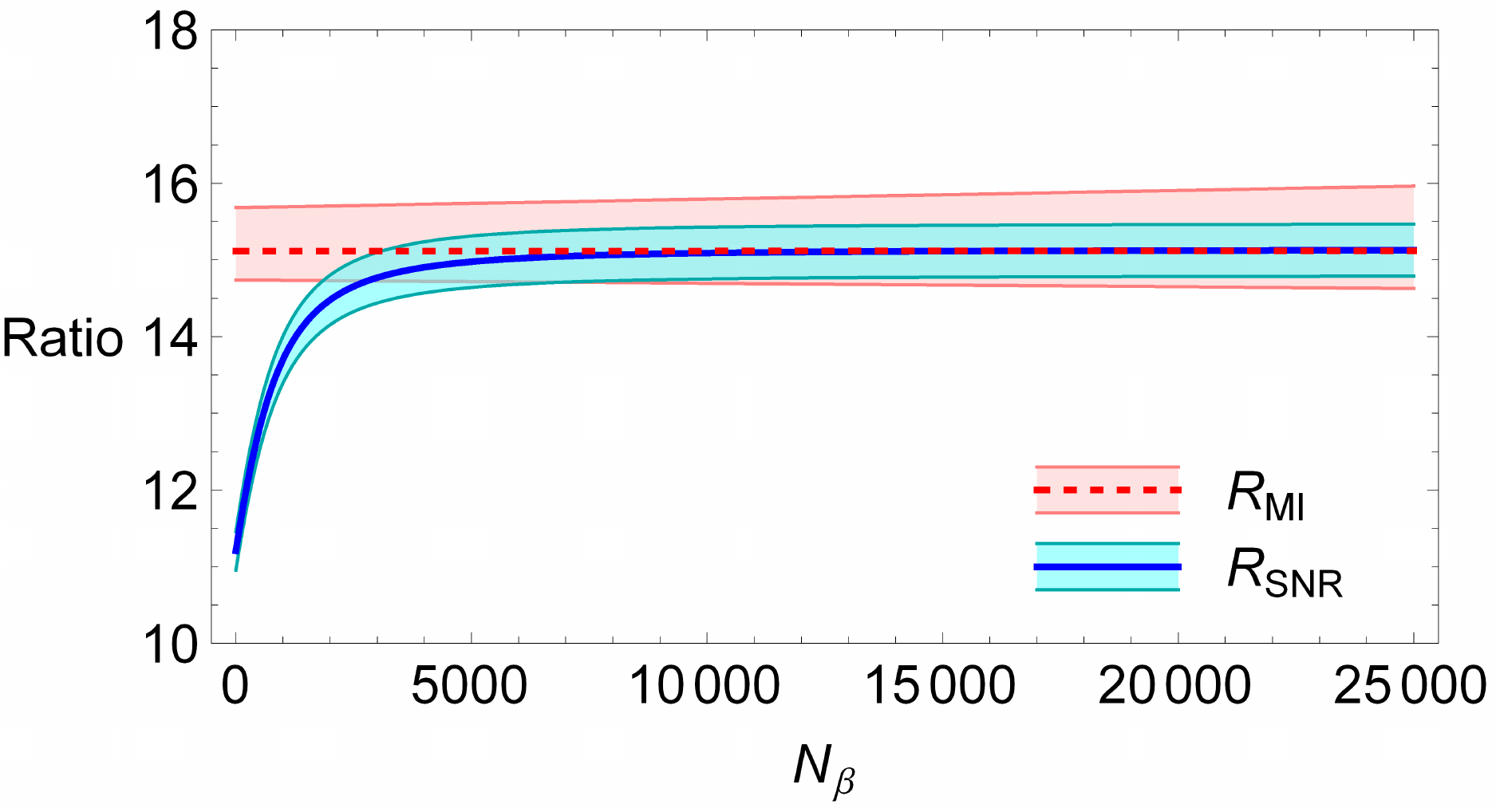}}\\
\caption{(a)--(d) Experimental results and theoretical expectations for the SNRs (blue solid) and effective MIs (red dashed) of the quantum illumination demonstration of Lopaeva {\it et al.} \cite{Lopaeva} using TWB and THB light, plotted versus the bath photon number $N_\beta$. The values of the other experimental parameters are fixed at  $M=90000$, $M_\beta=1300$, $\eta=0.38$, $\eta_\beta=0.5$, $N_{\text{TWB}}=4232$, $N_{\text{THB}}=3278$, for all the plots in this figure. (e) Ratios $\tilde R$ between SNR and MI for TWB (magenta empty squares) and THB (green filled triangles) sources, obtained from the measured data; the dashed gray curve represents the theoretical prediction, given by  the mean between $\tilde R_{\text{TWB}}$ and $\tilde R_{\text{THB}}$. (f) Confidence intervals (shadings) inferred from the experimental data, along with theoretical predictions (lines), for the ratios $R_{\text{SNR}}$ (blue solid) and $R_{\text{MI}}$ (red dashed).}
\label{PlotExp}
\end{figure}

\section{Comparisons with experiment}\label{s:exp}

The practical description introduced in the previous section allows us to directly assess the experimental results obtained in \cite{Lopaeva} . Assuming knowledge of the mode-count, we can extract our effective second-order correlations from intensity measurements and covariances. For example, $\langle\hat{a}^\dagger_{\text{eff}} \hat{a}_{\text{eff}} \rangle= \langle N/M \rangle$, from which the auto-correlations can be deduced with ease.
\par
In Fig.~\ref{PlotExp}(a)--(b) we plot the SNRs for TWB and THB as determined experimentally in \cite{Lopaeva}. In the same figure, panels (c)--(d), we plot the corresponding MIs obtained from the experimental data by constructing the effective two-mode operators as detailed above. We observe that a very good agreement is reached with the theoretical expectations based on Eqs.~(\ref{CMs}), especially in the case of TWB light. The THB case is affected by considerably lower accuracy: this is consistent with the intrinsic lower SNR of the measurement with THB (since $R\sim10$, we note that in order to achieve the same accuracy, a number of acquisitions 100 times larger than in the case of TWB illumination would be required for THB sources).
\par
According to the theoretical expectations, we should find that $$R_{\text{SNR}} \equiv \frac{\text{SNR}_\text{TWB}}{\text{SNR}_\text{THB}} \approx \frac{\text{MI}_\text{TWB}}{\text{MI}_\text{THB}} \equiv R_{\text{MI}},$$ or, equivalently, $$\tilde{R}_{\text{TWB}}\equiv\frac{\text{SNR}_\text{TWB}}{\text{MI}_\text{TWB}} \approx \frac{\text{SNR}_\text{THB}}{\text{MI}_\text{THB}} \equiv \tilde{R}_{\text{THB}},$$
in the relevant regime of high $N_\beta$. In Fig.~\ref{PlotExp}(e), we plot the ratios $\tilde{R}$ for TWB and THB light, respectively, as calculated directly from the measured data. We see indeed that the two quantities align with good precision along the same curve, in agreement with the theory.

Finally, to quantify the quantum enhancement in the implemented instance of continuous variable quantum illumination, we extrapolate the confidence intervals for the direct ratios $R_{\text{SNR}}$ and $R_{\text{MI}}$ and plot them in Fig.~\ref{PlotExp}~(f) against the theory (similarly to Fig.~\ref{PlotBath}). We conclude that the quantum enhancement allowed by the TWB over the corresponding THB with the same single-mode statistics is of a factor $\approx 15.1$, as determined by the asymptotic value of the ratios in the $N_\beta \gg 1$ regime, Eq.~(\ref{cross}). Note that we cannot directly calculate these ratios from the experimental points without resorting to model fitting, as the acquisitions in \cite{Lopaeva} correspond to different values of $N_{\beta}$ between the TWB and THB settings [see e.g.~Fig.~\ref{PlotExp}~(e)].

\section{Conclusion}\label{s:concl}

We have shown that for a continuous variable realization of quantum illumination \cite{Lloyd,Tan} as demonstrated experimentally in \cite{Lopaeva}, the fractional increase of mutual information (in an effective two-mode description) for entangled twin beams over correlated thermal beams provides a close approximation for the equivalent ratio of the SNR, which becomes exact in the practically relevant regime of a lossy system with a large number of thermal photons in the bath. This observation, as well as connecting to previous work on correlations in intensity interferometry setups \cite{Ghost,HBT}, provides insight into the source of quantum improvement in continuous variable quantum illumination. We neatly observed the predicted correspondence from the experimental data of Ref.~\cite{Lopaeva}.

We would like to point out that these results complement and do not controvert the results of \cite{vedral:13} or \cite{Giovannetti} for discrete variable quantum illumination. In particular, in \cite{vedral:13} the discord consumption, i.e.~the difference between the discord in the source light before and after the interaction with the target, is linked quantitatively to the quality of the protocol. Since some discord remains even when the initial entanglement is destroyed, the authors of \cite{vedral:13} conclude that discord plays a key role in empowering quantum rather than classical illumination. In the continuous variable setting, the Gaussian discord consumption is known to relate to a quantum advantage in a simple protocol of information encoding \cite{Gu:NP}, but such a scenario has not been investigated to date for the setting of quantum illumination (and it can be a good topic for further study).

Our paper offers an alternative perspective, where the correlations evaluated just for the source light are chosen as the object of study, in analogy to \cite{Ghost,HBT}. While clearly the mutual information includes both classical and quantum portions, we find  that in the coarse-grained two-mode description adopted here it is the total effective correlation, rather than just the effective discord of the source, that is found to capture the quantum advantage in a quantitative fashion. The central observation is that entangled states can be  overall more correlated (classically and quantumly) than separable states, for a given mean energy of the states. The resilience of these extra correlations, which we quantify via the mutual information in the effective picture, is here shown to capture the quantum enhancement, even when external noise degrades those correlations to the point that the quantum signature of entanglement is completely suppressed.

\subsection*{Acknowledgments}
This research was supported by the University of Nottingham through an EPSRC Research Development Fund Grant (PP-0313/36), the EU FP7
under grant agreement n.~308803 (BRISQ2), the Compagnia di San Paolo and MIUR (FIRB ``LiCHIS'' -- RBFR10YQ3H, Progetto Premiale ``Oltre i limiti classici di misura''). We thank S. Pirandola for discussions.


\clearpage
\end{document}